\documentclass[fleqn,usenatbib]{mnras}

\usepackage{newtxtext,newtxmath}

\usepackage[T1]{fontenc}

\DeclareRobustCommand{\VAN}[3]{#2}
\let\VANthebibliography\thebibliography
\def\thebibliography{\DeclareRobustCommand{\VAN}[3]{##3}\VANthebibliography}

\usepackage{graphicx}	
\usepackage{amsmath}	
\usepackage{amssymb}	

\usepackage{xspace}
\newcommand{\geu}{iPTF16geu\xspace}

\newcommand{\sn}{SN\xspace}
\newcommand{\snia}{SN~Ia\xspace}
\newcommand{\sneia}{SNe~Ia\xspace}
\newcommand{\sne}{SNe\xspace}
\newcommand{\hst}{HST\xspace}

\newcommand{\halpha}{H$_\alpha$\,}
\newcommand{\oii}{[\ion{O}{ii}]\,}
\newcommand{\nii}{[\ion{N}{ii}]\xspace}
\newcommand{\sii}{[\ion{S}{ii}]\,}

\newcommand{\kms}{km\,s$^{-1}$}
\newcommand{\kmsd}{km\,s$^{-1}$\,d$^{-1}$}

\title[Spectroscopy of iPTF16geu]{Spectroscopy of the first resolved strongly lensed Type Ia supernova iPTF16geu}

\author[Johansson et al.]{%
J.~Johansson,$^{1,2}$\thanks{E-mail: joeljo@fysik.su.se}
A.~Goobar,$^{2}$
S.~H.~Price,$^{3}$
A.~Sagu\'es~Carracedo,$^{2}$
L.~Della~Bruna,$^{4}$
\newauthor
P.~E.~Nugent,$^{5,6}$
S.~Dhawan,$^{2}$ 
E.~M\"ortsell,$^{2}$
S.~Papadogiannakis,$^{2}$
R.~Amanullah,$^{2}$
\newauthor
D.~Goldstein,$^{7}$
S.~B.~Cenko,$^{8,9}$
K.~De,$^{7}$
A.~Dugas,$^{7}$
M.~M.~Kasliwal,$^{7}$
S.~R.~Kulkarni,$^{7}$
\newauthor
R.~Lunnan$^{4}$
\\
$^{1}$Department of Physics and Astronomy, Division of Astronomy and Space Physics, Uppsala University, Box 516, SE 751 20 Uppsala, Sweden\\
$^{2}$The Oskar Klein Centre, Physics Department, Stockholm University, Albanova University Center, SE 106 91 Stockholm, Sweden\\
$^{3}$Max-Planck-Institut f\"ur Extraterrestrische Physik, Postfach 1312, Garching, 85741, Germany \\
$^{4}$The Oskar Klein Centre, Department of Astronomy, Stockholm University, AlbaNova University Centre, SE 106 91 Stockholm, Sweden\\
$^{5}$Lawrence Berkeley National Laboratory, 1 Cyclotron Road, Berkeley, CA, 94720, USA\\
$^{6}$Department of Astronomy, University of California, Berkeley, 94720, USA\\
$^{7}$Division of Physics, Mathematics and Astronomy, California Institute of Technology, Pasadena, CA 91125, USA\\
$^{8}$Astrophysics Science Division, NASA Goddard Space Flight Center, MC 661, Greenbelt, MD 20771, USA\\
$^{9}$Joint Space-Science Institute, University of Maryland, College Park, MD 20742, USA\\
}

\date{Accepted XXX. Received YYY; in original form ZZZ}

\pubyear{2015}

\begin{document}
\label{firstpage}
\pagerange{\pageref{firstpage}--\pageref{lastpage}}
\maketitle

\begin{abstract}
We report the results from spectroscopic observations of the multiple images of the strongly lensed Type Ia supernova (\snia), \geu, obtained with ground based telescopes and the {\em Hubble Space Telescope (HST)}. From a single epoch of slitless spectroscopy with \hst, we can resolve spectra of individual lensed supernova images for the first time. This allows us to perform an independent measurement of the time-delay between the two brightest images, $\Delta t = 1.4 \pm 5.0$ days, which is consistent with the time-delay measured from the light-curves.

We also present measurements of narrow emission and absorption lines characterizing the interstellar medium in the \snia host galaxy at $z=0.4087$, as well as in the foreground lensing galaxy at $z=0.2163$. 
We detect strong \ion{Na}{id} absorption in the host galaxy, indicating that \geu belongs to a subclass of \sneia displaying "anomalously" large \ion{Na}{id} column densities in comparison to the amount of dust extinction derived from their light curves. For the deflecting galaxy, we refine the measurement of the velocity dispersion, $\sigma = 129 \pm 4$ \kms, which significantly constrains the lens model. 

Since the time-delay between the SN images is negligible, we can use unresolved ground based spectroscopy, boosted by a factor $\sim 70$ from lensing magnification, to study the properties of a high-$z$ \snia with unprecedented signal-to-noise ratio. The spectral properties of the supernova, such as pseudo-Equivalent widths of several absorption features and velocities of the \ion{Si}{ii}-line indicate that \geu, besides being lensed, is a normal \snia, indistinguishable from well-studied ones in the local universe, providing support for the use of \sneia in precision cosmology. We do not detect any significant deviations of the SN spectral energy distribution from microlensing of the SN photosphere by stars and compact objects in the lensing galaxy. 

\end{abstract}



\section{Introduction}
More than half a century has passed since
\citet{1964MNRAS.128..307R} proposed to measure the expansion rate of the universe, the Hubble parameter, using time delay measurements of multiply-imaged gravitationally lensed supernovae (\sne). Thanks to the new generation of large time domain surveys, we are entering a phase where we can expect this technique to flourish 
\citep[][]{2002A&A...393...25G,Oguri2010,2017ApJ...834L...5G,Goldstein2019,Huber2019,suyu2020}. We refer to \citet{Oguri2020} for an excellent review of the field of lensing of explosive transients. 

The first case of a spatially resolved \snia is \geu \citep{Goobar2017}, a four-image gravitational lens system and the subject of this work. In accompanying papers \citep{dhawan2020,mortsell2019} we report on the photometric measurements of time-delays, extinction and magnification of \geu as well as the lens model. 
In addition to the time-delay estimates from photometric measurements, the well known temporal spectral evolution of \sneia \citep{hsiao2007} can also be used to constrain the time-delay between images, through spatially resolved spectroscopy. We have obtained \hst data for this purpose.

The high lensing amplification, $\mu \sim 70$, allows us to obtain very high signal-to-noise ratio spectroscopic observations that can be used to test the "standard candle" nature of \sneia, as done previously for PS1-10afx by \citet{2017A&A...603A.136P}. Furthermore, the amplification allows for spectroscopy with high spectral resolution, otherwise unfeasible for high-redshift SNe. These observations allow us also to scrutinize the host and lens galaxy properties. 
The large difference in the relative magnification between the four \sn images, in spite of the spatial symmetry, suggests that milli- or microlensing by substructures in the lensing galaxy is significant \citep[see e.g.,][]{2017ApJ...835L..25M,yahalomi2017,2018MNRAS.478.5081F,mortsell2019}. The impact of microlensing of strongly lensed \sne has been quantified by \citet{2006ApJ...653.1391D} and, more recently, \citet{2018ApJ...855...22G, Huber2019,2019ApJ...876..107P} carried out simulations including the effect of an expanding \sn photosphere to show that microlensing often induces chromatic effects on the supernova spectrum, most noticeable starting about a month after the explosion. Here we explore this possibility using a time series of \geu spectra.

The paper is organized as follows: First, in Sect.~\ref{sec:observations} we describe the photometric and spectroscopic data that we will use in our analysis. Then, in Section~\ref{sec:lenshost} we construct a model spectral energy distribution (SED) of the lens and host galaxy, that we can subtract from our observed spectra. We also analyse emission and absorption lines in our highest resolution spectra, probing the velocity dispersion of the lens and the interstellar medium of the host galaxy of \geu. In Sect.~\ref{sec:snfeatures} we analyse the time-evolution of the SN features, measuring their pseudo-equivalent widths and expansion velocities. In Sect.~\ref{sec:timedelay} we analyse our single epoch of slitless spectroscopy with HST, and perform an independent measurement of the time-delay between the brightest SN images. We conclude with a discussion, summary and future outlook in Sect. \ref{sec:discussion} and \ref{sec:summary}.

\section{Observations}\label{sec:observations}
The main focus of this paper is the analysis of the ground- and space-based spectroscopic observations of iPTF16geu. In order to accurately account for the lens and host galaxy contribution to the observed SN spectra we will use the \hst and ground-based photometry described in \citet{Goobar2017} and \citet{dhawan2020}.

\begin{figure}
	\centering
	\includegraphics[width=\columnwidth]{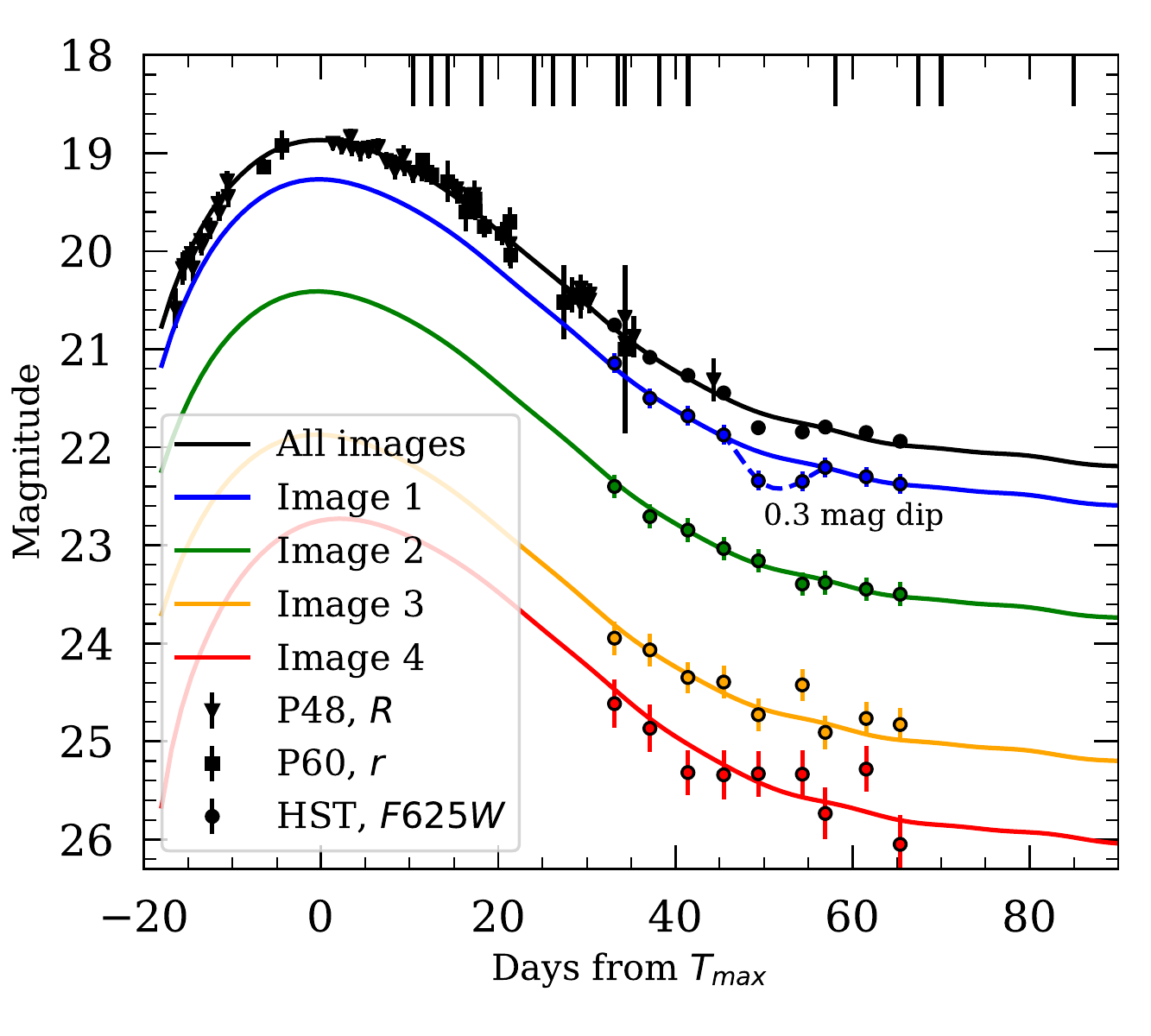}
	\caption{
	\hst $F625W$ and ground based $r$-band lightcurve of \geu (all four images summed). The solid lines are fits to the summed and individual light-curves, and the vertical black lines indicate when the spectra were observed. We note a small dip ($\sim 0.3$ mag) in the lightcurve of Image 1, not seen for the other SN images nor in the $F475W$ and $F814W$ data.
	\label{fig:lc}}
\end{figure}

\begin{figure}
	\centering
	\includegraphics[width=\columnwidth]{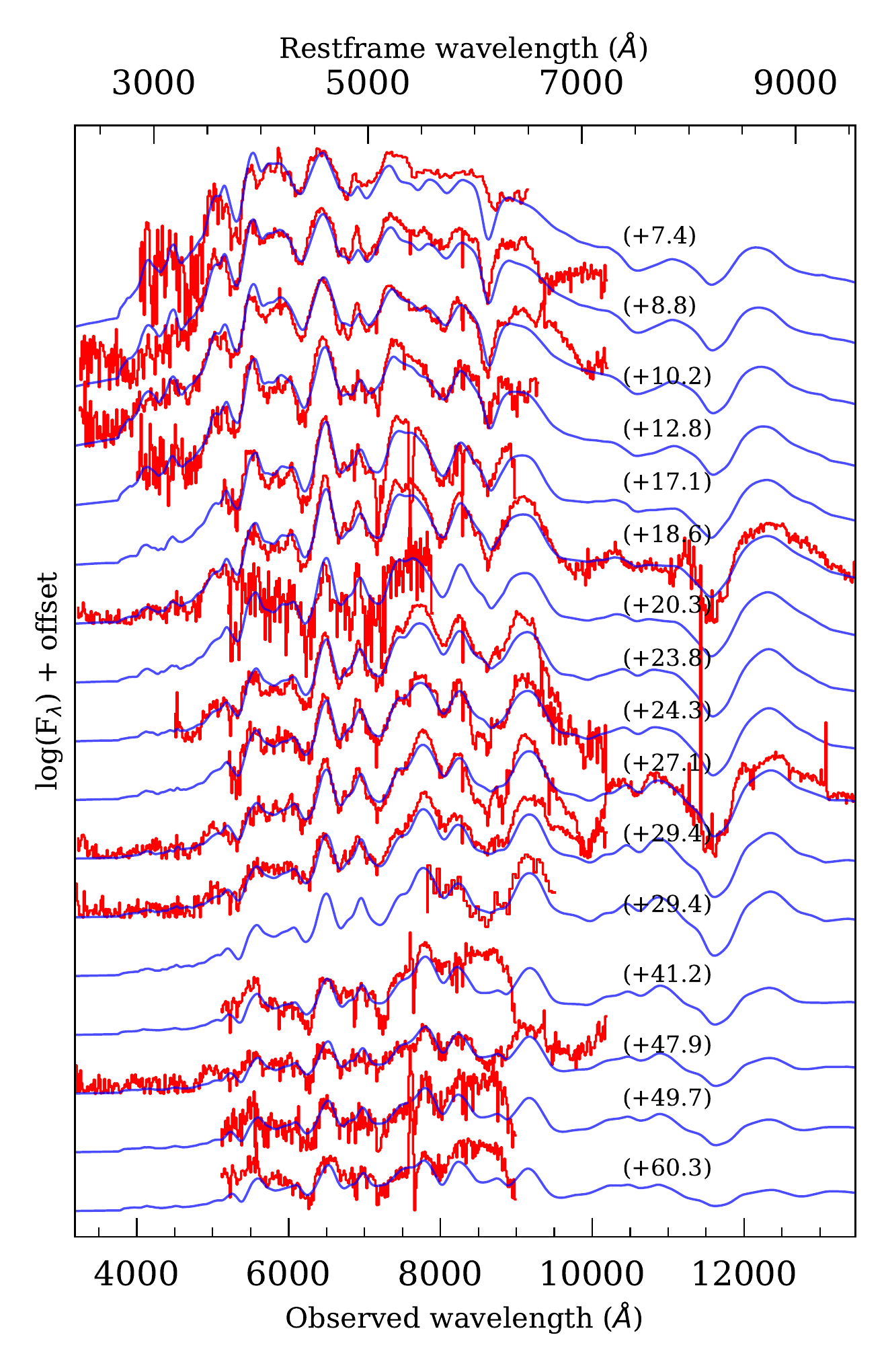}
	\caption{
	Time series of all spectra of iPTF16geu. The spectra (red lines) have been corrected for lens and host galaxy contamination. Blue lines are reddened Hsiao template SEDs at the same epochs.
	\label{fig:all_spectra}}
\end{figure}

Figures~\ref{fig:lc} and \ref{fig:all_spectra} illustrate the observations used in this work. Figure~\ref{fig:lc} shows the ground-based P48 $R$, P60 $r$-band and \hst $F625W$ light-curves of the four SN images (summed lightcurves in black, and resolved photometry for Images 1,2,3,4 in blue, green, orange and red). The epochs of spectroscopic observations are indicated by black, vertical lines.

\subsection{Photometry}
\begin{figure*}
	\centering
	\includegraphics[width=\textwidth]{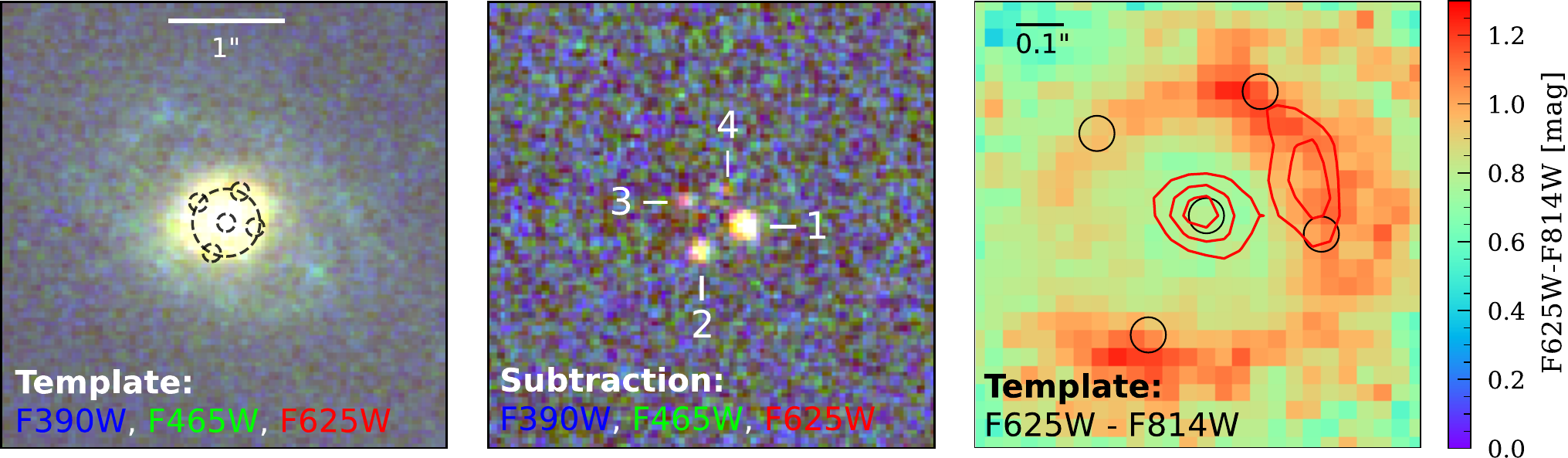}
	\caption{Left panel: \hst template images of the lens and host galaxies obtained after the SN faded. We note that the lens galaxy has a faint, diffuse blue halo. Middle panel: A difference image between the template and SN images (from 2016-10-25). Right panel shows the $F625W-F814W$ color of the lens and host galaxy system from template \hst images (zoomed in on the central $1" \times 1"$). The red contour lines show the brightness distribution in $F814W$, indicating that the bright lens galaxy center has an observed color $F625W-F814W \approx 0.8$ mag, whereas the lensed host galaxy ring has a color that varies between $F625W-F814W \approx 0.9-1.3$ mag. \label{fig:colorimage}}
\end{figure*}

\citet{dhawan2020} present \hst photometry of the four resolved SN images, and fit the individual light-curves together with the summed photometry from ground based telescopes. For Image 1 (the brightest image, see Fig.~\ref{fig:colorimage}), they find $T_{\rm max,1} = 57652.80 \pm 0.33$  (MJD, which we will refer to as the time of maximum light) and a stretch $s=0.99 \pm 0.01$. The measured time delays between the four images are consistent with being less than 1 day and the four lines of sight through the host and lensing galaxy experience differential extinction. Furthermore, the lensing analyses in \citet{2017ApJ...835L..25M} and \citet{mortsell2019} indicate that \geu, especially the brightest image (Image 1), is likely affected by additional "micro-lensing" from sub-structures in the lens galaxy halo.

We make use of the resolved \hst and ground-based SN photometry presented in \citet{Goobar2017} and \citet{dhawan2020}. In Sec.~\ref{sec:lenshost}, we also use of pre- and post-SN photometry of the lens and host galaxy system from the SDSS, PS1 and 2MASS surveys as well as the \hst reference images.

\subsection{Spectroscopy}

\begin{table}
	\centering
	\caption{Log of spectroscopic observations of \geu used in this analysis.}
		\label{tab:spectra_log}
\begin{tabular}{llllc@{}}
\hline\hline
UT date & MJD   & Phase & Telescope+Instrument & Source \\
& & (days) & & \\
\hline
2016-10-02.23 & 57663.23 & +7.4 & P60+SEDM & 1\\ 
2016-10-04.22 & 57665.22 & +8.8 & P200+DBSP & 1 \\ 
2016-10-06.13 & 57667.13 & +10.2 & P200+DBSP & 1 \\
2016-10-09.90 & 57670.90 & +12.8 & NOT+ALFOSC  & 1 \\ 
2016-10-15.87 & 57676.87 & +17.1  & GTC+OSIRIS & 2 \\ 
2016-10-18.00 & 57679.00 & +18.6  & VLT+XSHOOTER & 2 \\ 
2016-10-22.34 & 57681.34 & +20.3 & DCT+DeVeny & 3 \\
2016-10-25.33 & 57686.33 & +23.8 & Keck+DEIMOS & 3 \\ 
2016-10-26.19 & 57687.10 & +24.3 & P200+DBSP & 3\\ 
2016-10-30.01 & 57691.01 & +27.1 & VLT+XSHOOTER & 2 \\ 
2016-11-02.24 & 57694.24 & +29.4 & Keck+LRIS & 3 \\ 
2016-11-02.27 & 57694.27 & +29.4 & HST+WFC3/IR & 3 \\
2016-11-18.86 & 57710.86 & +41.2 & GTC+OSIRIS & 2 \\
2016-11-28.21 & 57720.21 & +47.9 & Keck+LRIS & 3 \\ 
2016-11-30.82 & 57722.82 & +49.7 & GTC+OSIRIS & 2 \\
2016-12-15.81 & 57737.81 & +60.3 & GTC+OSIRIS & 2 \\ 
\hline
2019-05-24.46 & 58627.46 & - & P200+DBSP & 3 \\ 
2019-06-26.46 & 58660.46 & - & P200+DBSP & 3 \\ 
\hline\hline \\
\multicolumn{5}{l}{1: \citet{Goobar2017}, 2: \citet{Cano2018}, 3: This work}\\
\end{tabular}

\end{table}

A summary of our spectroscopic observations are listed in Table~\ref{tab:spectra_log} and the time series of spectra are shown in Fig.~\ref{fig:all_spectra}. In addition to the previously unpublished spectra presented here, we also analyse the early spectra from \citet{Goobar2017} and the GTC+OSIRIS and VLT+XSHOOTER spectra from \citet{Cano2018}. All our spectra are available from the WISeREP archive \footnote{http://wiserep.weizmann.ac.il} \citep{2012PASP..124..668Y}.

The long-slit spectra presented here were reduced in a standard fashion using \texttt{IRAF}\footnote{The Image Reduction and Analysis Facility is distributed by the National Optical Astronomy Observatory, which is operated by the Association of Universities for Research in Astronomy (AURA) under cooperative agreement with the National Science Foundation (NSF).} routines. To reduce and extract the P200+DBSP and Keck+LRIS spectra we use the custom  \texttt{pyraf-dbsp} and \texttt{LPipe} pipelines \citep{Bellm2016,LPipe}. All ground based spectra were observed using a slit width of $\sim 1 - 2$", covering all four SN images including the lens and host galaxies. The spectra were observed at parallactic angle, except the two VLT+XSHOOTER spectra, which were observed at $PA=+86 \deg$. 
We note that the GTC+OSIRIS spectra from \citet{Cano2018} have not been corrected for telluric lines and suffer from fringing and second order contamination above 8300 \AA.

The VLT+XSHOOTER spectra were observed using an ABAB nod-on-slit pattern. The ESO XSHOOTER pipeline is used for bias subtraction, flatfield correction, wavelength and flux calibration using spectrophotometric standards, and sky-subtraction of the spectra taken at each position of the slit. We use pre- and post-processing scripts from \citet{selsing2019} \footnote{https://github.com/jselsing/xsh-postproc} to remove cosmics, extract and combine the 1D spectra. Telluric lines are corrected using Molecfit and the parameter file provided by ESO.

We also obtained one epoch of slitless spectroscopy using HST WFC3/IR GRISM256/G102. The grism spectra were reduced using \texttt{GRIZLI} (see \citealt{Wang18}, \citealt{Brammer19}\footnote{Using v0.2.1-39, from https://github.com/gbrammer/grizli}), a tool to reduce, model, and extract slitless spectra. \texttt{GRIZLI} first performs raw processing of direct F105W and G102 exposures, makes a composite F105W image, and constructs a source catalog and segmentation map \citep[with various individual internal steps performed with AstroDrizzle and Source Extractor;][]{Bertin96}. Individual grism exposures are then corrected for contamination from the other sources using iterative contamination modeling, using the masked F105W image of each source as the spatial profile (e.g., as in \citealt{Abramson19}). 
 
\begin{figure*}
	\centering
	\includegraphics[width=\textwidth]{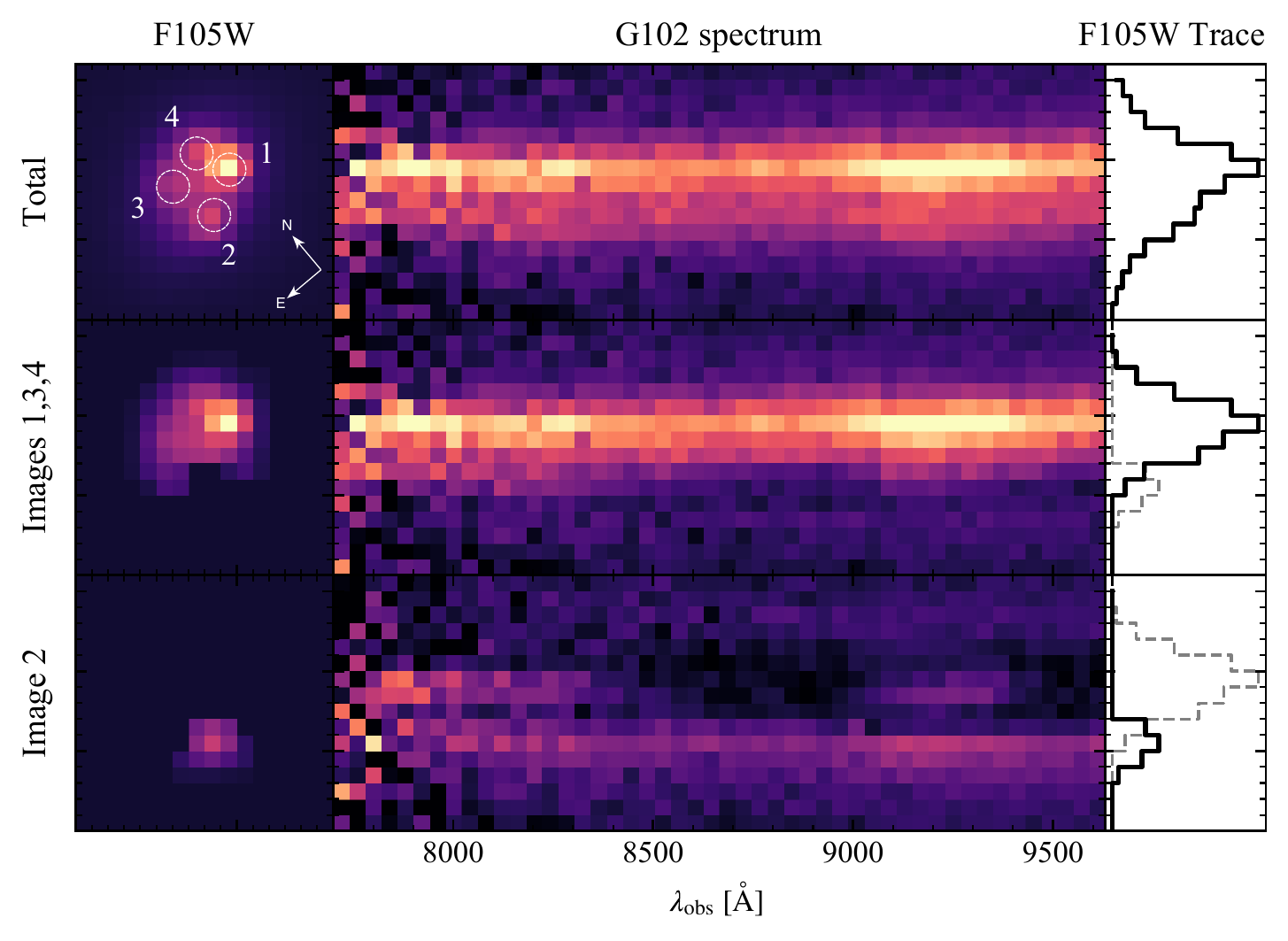}
	\caption{
    HST $F105W$ direct imaging (left column), resolved HST $G102$ grism spectra (center), and collapsed F105W trace used for optimal 1D extraction (right) for the lens and host galaxies (top row). Grism spectra for Images 1, 3, and 4 (middle row) and Image 2 (bottom) are separately extracted. The 1D spectra of the image groupings are extracted from the contamination-subtracted grism spectra (center) using the masked $F105W$ trace (right) as the optimal extraction  profile (solid black line; other trace shown as dashed grey line). 
    \label{fig:hstextraction}}
\end{figure*}

Composite 2D grism spectra for each source are then constructed from these contamination-corrected individual exposures (see Fig.~\ref{fig:hstextraction}). Finally, we extract optimally-weighted 1D spectra \citep{Horne86} for each source from the composite 2D spectra, adopting the collapsed composite masked F105W image trace as the optimal spatial profile. 
In this analysis, the F105W image and G102 grism spectra are split into two parts, where we extract two separate spectra: the first corresponding to a blend of Images 1, 3, and 4 (Image 1 being the dominant), and a second being Image 2 (the second brightest, south east image).

\section{The iPTF16geu lens and host galaxies}\label{sec:lenshost}

\begin{figure}
	\centering
	\includegraphics[width=\columnwidth]{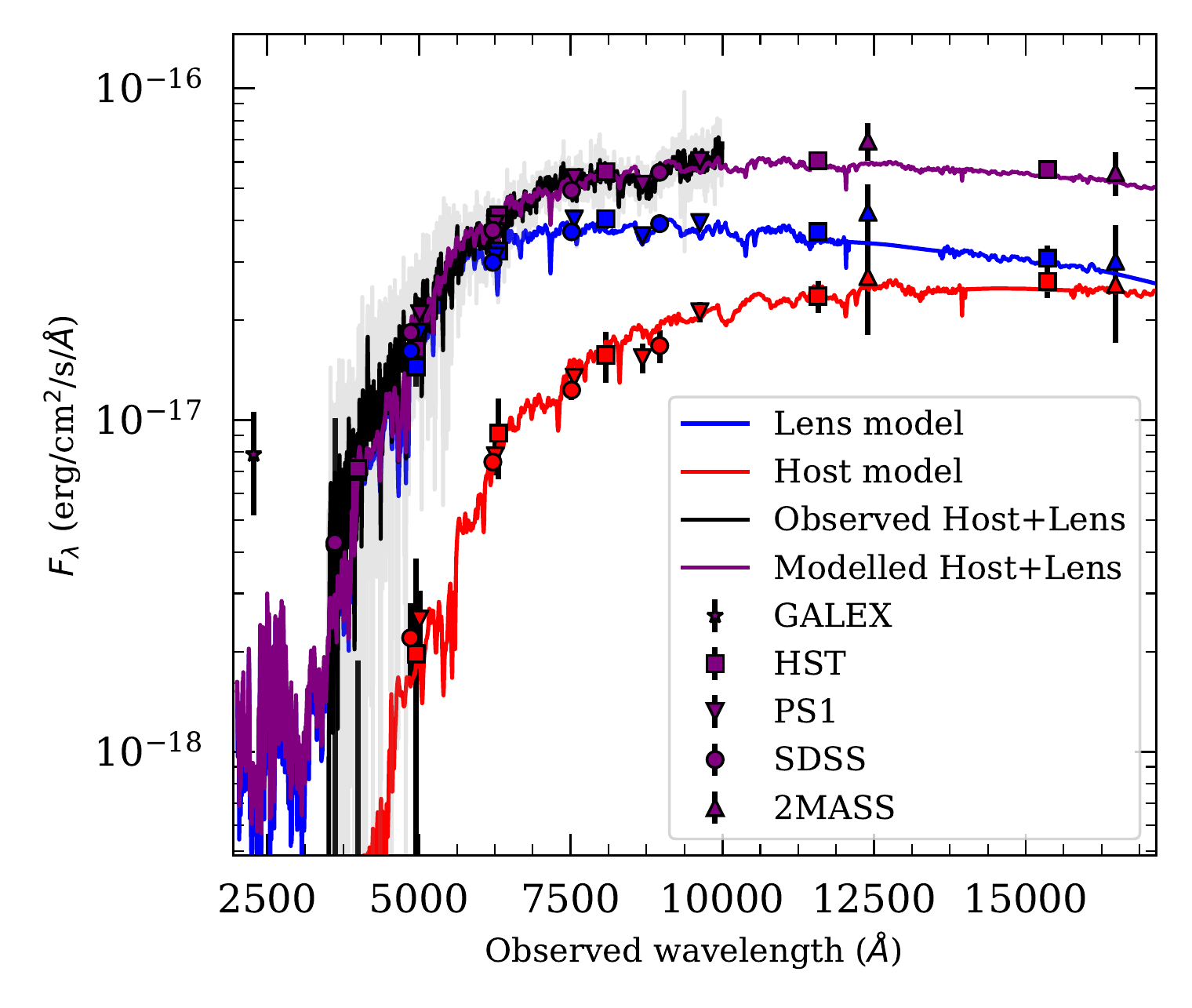}
	\caption{SED of lens and host galaxies. Purple symbols show broad-band GALEX, SDSS ($ugriz$), Pan-Starrs ($grizy$), 2MASS ($JHK$) and HST photometry of the lens and host galaxies. The blue line is the lensing galaxy at $z=0.2163$ (reddened by galactic extinction, $E(B-V)_{\rm MW}=0.073$ mag), the red line is the host galaxy at $z=0.4087$ (reddened by extinction in the MW and the lensing galaxy, $E(B-V)_{\rm lens}=0.30$ mag). The purple line is the sum of both galaxy template spectra.
	The black line is a P200+DBSP spectrum, observed in 2019 after the SN faded. 
	\label{fig:lens_host_SED}}
\end{figure}

Unlike the case of e.g. strongly lensed quasars, we are able to separately observe the lens and host galaxy system before and after the SN was active. Besides allowing us to study the lens and host galaxy properties, we can use pre- and post SN data to remove the lens and host galaxy contribution from the observed SN spectra.

Two years after the SN explosion, we obtained a low signal-to-noise spectrum of the lens and host galaxy system combining two epochs of P200+DBSP observations with 900 and 1800 seconds exposure time, covering 3500 - 10000 \AA\, (black line in Fig.~\ref{fig:lens_host_SED}). To further disentangle the contributions from the lens and host galaxies to the total flux over a wider wavelength range, we construct a model spectral template of the combined lens and host galaxies. To do this, we use broad band photometric data of the lens+host system from the SDSS ($ugriz$ filters), Pan-Starrs ($grizy$ filters) and 2MASS ($JHK$ filters) surveys predating the SN, as well as our \hst template images ($F390W,F475W,F625W,F814W,F110W$ and $F160W$ filters, observed after the SN faded). We then use the models of the lens and host galaxy light in \citet{dhawan2020} to compute the fractional fluxes of the lens and host galaxies (blue and red symbols in Fig.~\ref{fig:lens_host_SED}, respectively) within a 1" diameter aperture for all filters. We then fit for a combination of two galaxy template SEDs (Elliptical/S0/Sa/Sb or Sc) from \citet{mannucci2001} at the lens and host redshifts. Fig.~\ref{fig:lens_host_SED} shows the best-fit lens and host galaxy SED (blue and red lines in Fig.~\ref{fig:lens_host_SED}, respectively) and the total lens+host SED (purple symbols and line).

While we have little information on the morphology of the lens and host galaxy, we note that the fitted S\'ersic indices of the central lens galaxy core (within 0.3") from \hst and Keck AO images range between 0.8 - 1.6, which is consistent with that of an Elliptical or a Bulge galaxy. As seen in Fig.~\ref{fig:colorimage}, there is a faint extended blue halo (possibly spiral arms) around the bright lens galaxy core, and weak emission lines from $H_\alpha$, [\ion{N}{ii}] and [\ion{O}{ii}] at the lens rest-frame on top of broader absorption features. Similarly, weak emission lines from $H_\alpha$, [\ion{N}{ii}] and [\ion{O}{ii}] are also seen at the host redshift.

\citet{dhawan2020} used the SN images to measure the differential extinction along the line-of-sight towards the four point-like SN images. To verify the assumptions on the global dust extinction of the host galaxy by the lens, we investigate the color of the lensed host galaxy ring. The $F625W-F814W$ color of a Ellipctical (Sa) galaxy template is 0.89 (0.81) mag. Along the host galaxy ring, we measure $F625W-F814W$ = 0.9-1.3 mag, while the brightest host galaxy spot has a color of $F625W-F814W \approx 1.1$ mag (right panel in Fig.~\ref{fig:colorimage}). This motivates us to redden the host galaxy SED by $E(B-V) \approx 0.3$ mag.

\subsection{Redshift and velocity dispersion of the lens}
In the SN and galaxy spectra we see narrow emission lines from the \oii $\lambda3727$ doublet, \halpha, \nii (and weak emission from of \sii) at the lens galaxy redshift, indicating a low  level of star formation.

\citet{Goobar2017} analyzed the emission lines visible in the early, low-resolution, SN spectra and measured the lens galaxy redshift ($z_{\rm lens}=0.216$), and gave a first estimate of line-of-sight velocity dispersion, $\sigma_v = 163^{+41}_{-27}$ \kms, of the lensing galaxy from the combined widths of the \halpha and \nii lines.

\begin{figure*}
	\centering
	\includegraphics[width=\textwidth]{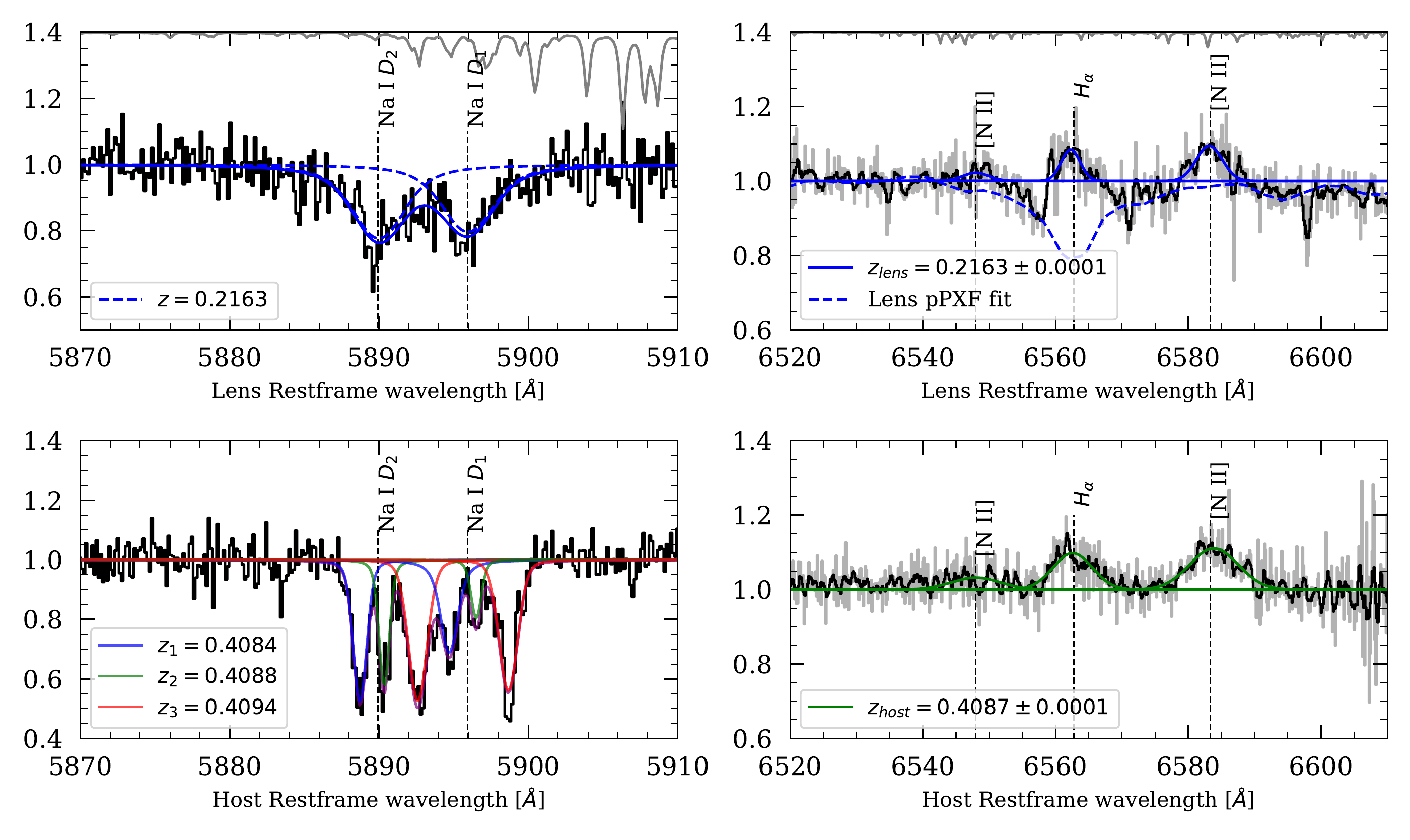}
	\caption{Zoomed in panels on narrow \ion{Na}{id} absorption lines, H$_{\alpha}$ and [\ion{N}{ii}] emission lines at the lens ($z_{\rm lens}=0.2163$) and host ($z_{\rm host}=0.4087$) rest-frames (marked by vertical black dashed lines in the panels) in the VLT+XSHOOTER spectrum from 2016 Oct. 30 (+27 days after maximum light). The grey line at the top of the upper panels shows the telluric lines, which have been removed from the spectra.
	\label{fig:abs_em_panels}}
\end{figure*}

Here, we use our highest resolution spectra (the two VLT+XSHOOTER spectra) and fit gaussian profiles to the emission lines. We do this after normalizing the spectra, fitting low-order polynomials to the continuum level around the regions of interest. The most prominent emission lines are shown in Fig.~\ref{fig:abs_em_panels} for the lens galaxy (top panels) and host galaxy (bottom panels). From these lines we determine the redshift of the lens, $z_{\rm lens} = 0.2163 \pm 0.0001$.

In the VLT+XSHOOTER spectra, we note that the lens galaxy \halpha emission line is affected by Balmer absorption from stellar atmospheres (seen from extrapolating the stellar continuum fit, shown as a blue dashed line in the upper right panel of Fig.~\ref{fig:abs_em_panels}) and hence the true strength is underestimated. 

\begin{figure}
	\centering
	\includegraphics[width=\columnwidth]{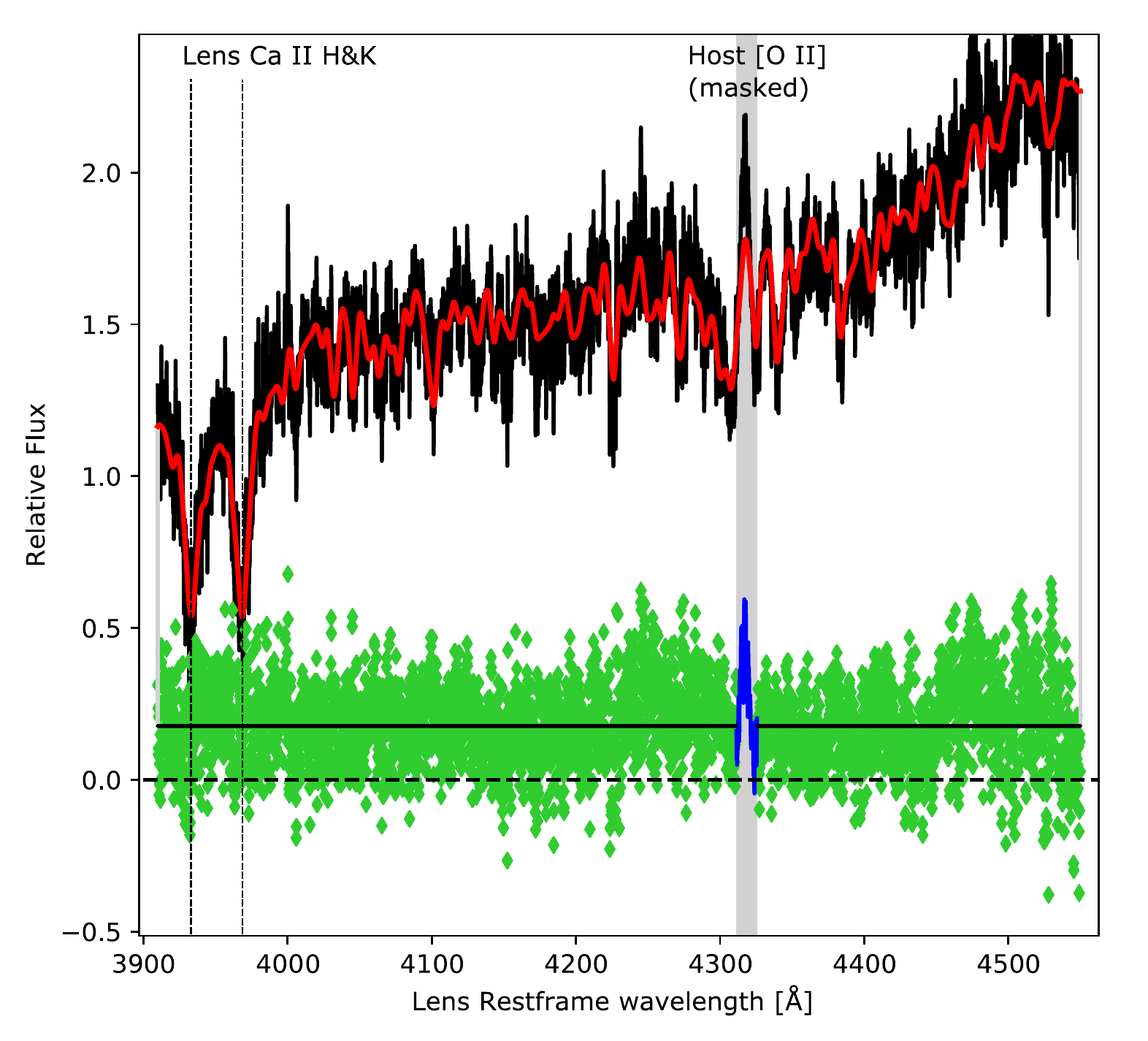}
	\caption{pPXF fit of the lens galaxy. The black line is part of the UVB-arm XSHOOTER spectrum from 2016 Oct. 30, red line is the best fit PEGASE template and green symbols show the residuals. The dashed lines indicate the lens galaxy \ion{Ca}{ii} H \& K absorption lines and the gray shaded area shows the masked-out [\ion{O}{ii}] line from the host galaxy.
	\label{fig:ppxf_fit}}
\end{figure}

For the lensing galaxy, we measure the equivalent widths (EW) of the \ion{Ca}{ii} H\&K and \ion{Na}{id} absorption features to be 11.5 \AA \, and 2.3 \AA \,, respectively. Both features are centered at the lens rest-frame redshift (see Figures \ref{fig:abs_em_panels} and \ref{fig:ppxf_fit}).

In order to estimate the velocity dispersion of the lens galaxy, we perform stellar continuum fitting with \textsc{pPXF} \citep{cappellari04,cappellari17}. \textsc{pPXF} models a galaxy spectrum $G$ as a convolution between template spectra $T$ and the line of sight velocity distribution (LOSVD) of the stars $\mathcal{L}$:
$$G_{mod} (x) = T(x) * \mathcal{L} (c x),$$
where $x = \ln \lambda$.

\par For the VLT+XSHOOTER ($R\approx 5400$) and Keck+DEIMOS ($R\approx 2000$) spectra, we use the high resolution PEGASE models~\citep{leborgne04}, spanning the wavelength range $3900-6800$~\AA\, at a FWHM $\sim 0.55$ \AA \ and having ages and metallicities in the range $t = 1 - 2 \times 10^4$ Myr and $Z = 0.0004 - 0.1$.
We mask the \oii line from the host, and use the restframe wavelength range 3900-4550 \AA\,(to avoid \ion{Ca}{ii} H\&K from the host) in the XSHOOTER spectrum, giving a best-fit velocity dispersion $\sigma = 129 \pm 4$ \kms. The data and fit are shown in Fig.~\ref{fig:ppxf_fit}. This measured velocity dispersion is lower than the previous estimate \citep{Goobar2017}, and matches the expected velocity dispersion $\sigma_{\rm mod}=132^{+4}_{-7}$ \kms from the lens modelling in \citet{mortsell2019}. 

\par As a consistency check, we also fit our low-resolution Keck+LRIS and post-SN P200+DBSP spectra using the UV-extended MILES templates \citep[eMILES;][]{vazdekis16}. These single-age, single-metallicity stellar population spectra and have a resolution of $2.51$~\AA\,(FWHM) in the range $3540 - 8950$~\AA. 
These spectra yield similar best-fit values of the velocity dispersion, but with larger errorbars ($\sim 25$ \kms).

\subsection{The host galaxy of iPTF16geu}
Thanks to the high lensing magnification we are able to study the structure and time evolution of interstellar absorption lines for \sneia at cosmological distances, at a level previously only possible a small subset of very nearby \sneia \citep[e.g.][]{goobar2014,ferretti2016,ferretti2017}. 

Similar to the lens galaxy, we also detect narrow emission lines from \oii $\lambda3727$, \halpha, \nii at the host galaxy redshift. From these lines we determine the and host galaxy redshift, $z_{\rm host} = 0.4087 \pm 0.0001$ (see right, bottom panel of Fig.~\ref{fig:abs_em_panels}).

For \ion{Na}{i}~D, we resolve three distinct, narrow components (FWHM $\sim 44-79$ \kms, see left, bottom panel of Fig.~\ref{fig:abs_em_panels}) at $v_1=-61$, $v_2=23$, $v_3=139$ \kms, with respect to the galaxy rest-frame (defined by the emission lines). For the first XSHOOTER epoch at +18.6 days we measure a total \ion{Na}{id} $EW=3.9$\, \AA, and for the second epoch $EW=3.3$\,\AA\,at +27.1 days. The EWs are listed in Table~\ref{tab:pew_vel}.

The equivalent width of the \ion{Na}{id} lines is a commonly used proxy for dust reddening, $E(B-V)$ \citep[see e.g.][]{munari1997,Poznanski2011,Poznanski2012}.
It should be noted that these relations are typically well-defined for $EW < 1.0$\,\AA \, and that the theoretical relation is between EW and optical depth, $\tau \sim A_V$, rather than the color excess $E(B-V)$. Nonetheless, applying these relations using and $EW^{\ion{Na}{id}}_{\rm host}=3.9$\,\AA\, yields $E(B-V)_{\rm host} \sim 0.6-1.9$ mag. 
The light-curve fits \citep{dhawan2020} indicate that the host extinction is $E(B-V)_{\rm host} = 0.17 - 0.29$ mag (depending whether the total-to-selective absorption ratio is fixed to $R_V$ = 2.0, 3.1 or as a treated as a free parameter). \geu therefore seems to be another case of a \snia displaying "anomalously" strong \ion{Na}{id} absorption \citep{phillips2013}.

We do not detect any absorption lines from Diffuse Interstellar Bands (e.g. DIB $\lambda$5780, which has been suggested to be an even better proxy for dust extinction).

\section{Supernova features and their time-evolution}\label{sec:snfeatures}
\begin{figure}
	\centering
	\includegraphics[width=\columnwidth]{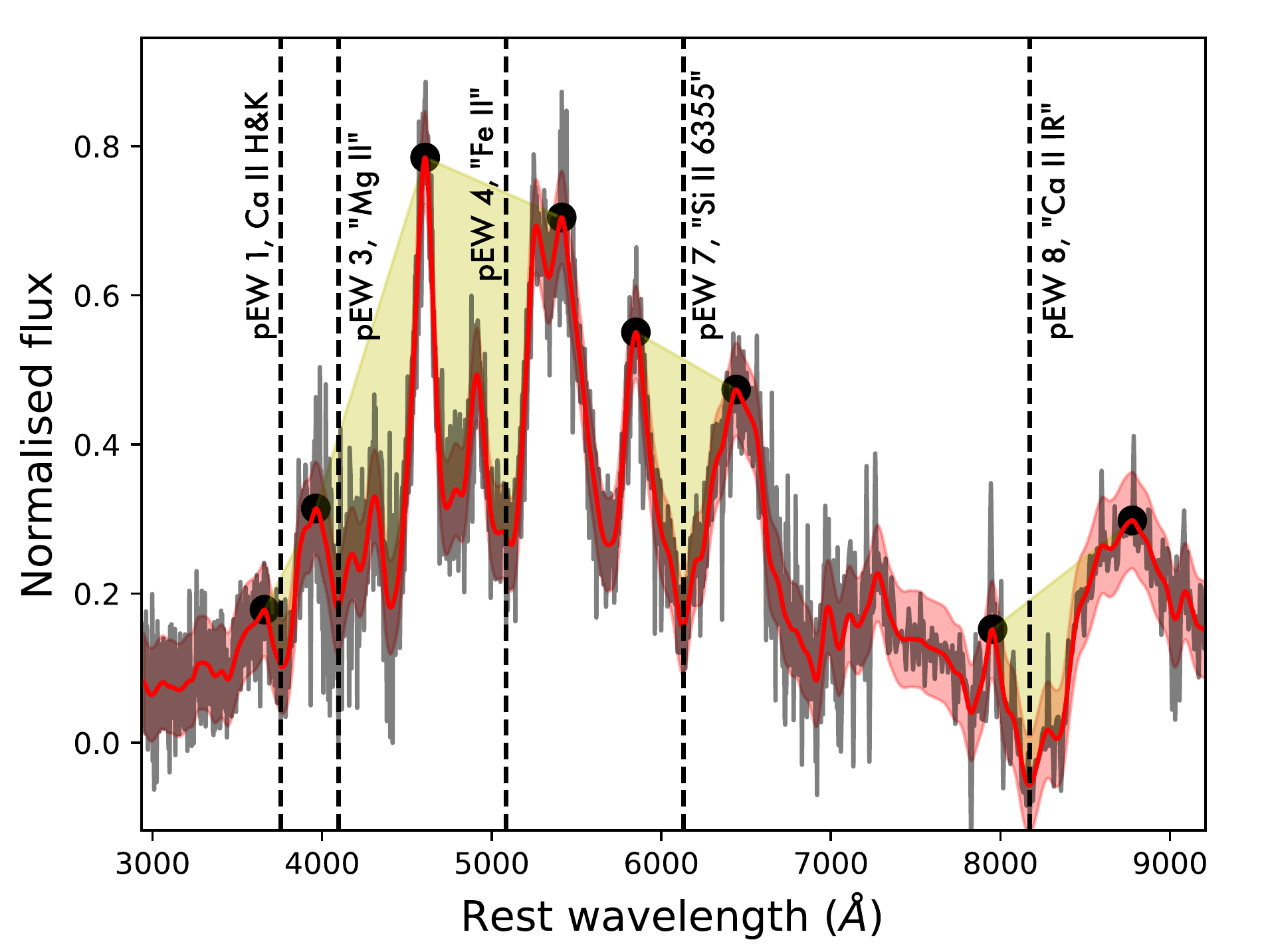}
	\caption{Example of pseudo-Equivalent Width measurements using spextractor. The black line is a lens and host subtracted VLT+XSHOOTER spectrum from 18.6 days after maximum light, the red line is the Gaussian process smoothed spectrum, the yellow regions show the measured pEWs and the black dashed lines indicate the fitted absorption line minima.
	\label{fig:spextractor}}
\end{figure}

\citet{Cano2018} found that \geu can be classified as a high-velocity ($v_{{\rm Si\,\sc{II}} \lambda6355}^{\rm max} = 11950 \pm 140$ \kms), high-velocity gradient ($\dot{v} = -110.3 \pm 10.0$ \kms) and "core-normal" SN Ia. The strength of various features (measured though their pseudo-equivalent widths) argue against SN \geu being a faint, broad- lined, cool or shallow-silicon SN Ia.

Using more data, and a refined lens and host galaxy template spectrum, we will measure the pseudo-equivalent widths (pEWs) and \ion{Si}{ii} line expansion velocities. As demonstrated in \citet{Cano2018}, accurate removal of the lens- and host galaxy contribution is crucial in order to measure the intrinsic SN pEWs. For our purposes, we subtract both our observed and model template spectrum from the observations, where the former allows us to measure pEW1 and pEW8. We first subtract the lens and host galaxy contamination from the observed spectra, scaling the observed spectra with a constant factor, so that the subtracted spectra match the template subtracted photometry (this scaling typically varies between 0.8 - 1.2, accounting for imperfect flux calibration, varying degrees of host galaxy removal in the different reduction procedures, etc.). While the spectra could in practice also suffer from wavelength-dependent calibration offsets (due to slit-losses, atmospheric dispersion etc.), we do not attempt to correct for this, since this might erase any chromatic micro-lensing signatures. Furthermore, we de-redden the summed spectra with the best fit lens and host galaxy extinction parameters found by \citet{dhawan2020}, using $E(B-V)^{\rm host} = 0.18$ mag with $R_{V}^{\rm host}=2.0$, $E(B-V)^{\rm lens,1} = 0.26$ mag with $R_{V}^{\rm lens, 1} = 1.8$ (since the ground-based spectra are dominated by the light from Image 1) and $E(B-V)_{MW} = 0.073$ mag.

To compute the pseudo-equivalent widths (pEWs) and \ion{Si}{ii} line expansion velocities, we use the \verb,spextractor, code (Papadogiannakis et al., in prep. \footnote{Code is publicly available at \url{github.com/astrobarn/spextractor}}). 
Instead of fitting a series of gaussians to the absorption features, spextractor measure the pEWs and absorption minima through model-independent gaussian processes.

In our spectra, ranging between +7 to +60 days from peak brightness, we can measure pEW1 (\ion{Ca}{ii}~H\&K), pEW3 (\ion{Mg}{ii}), pEW4 (\ion{Fe}{ii}), pEW7 (\ion{Si}{ii}~6355) and pEW8 (\ion{Ca}{ii} IR) \citep[following the conventions in][]{garavini2007,folatelli2004,folatelli2013}. Figure~\ref{fig:spextractor} shows an example of a lens and host galaxy subtracted, de-reddened VLT+XSHOOTER spectrum from 2016 Oct. 18. It includes the smoothed spectrum (red line), the pEW features and absorption line minima fitted by \verb,spextractor,.

\begin{figure*}
	\centering
	\includegraphics[width=\textwidth]{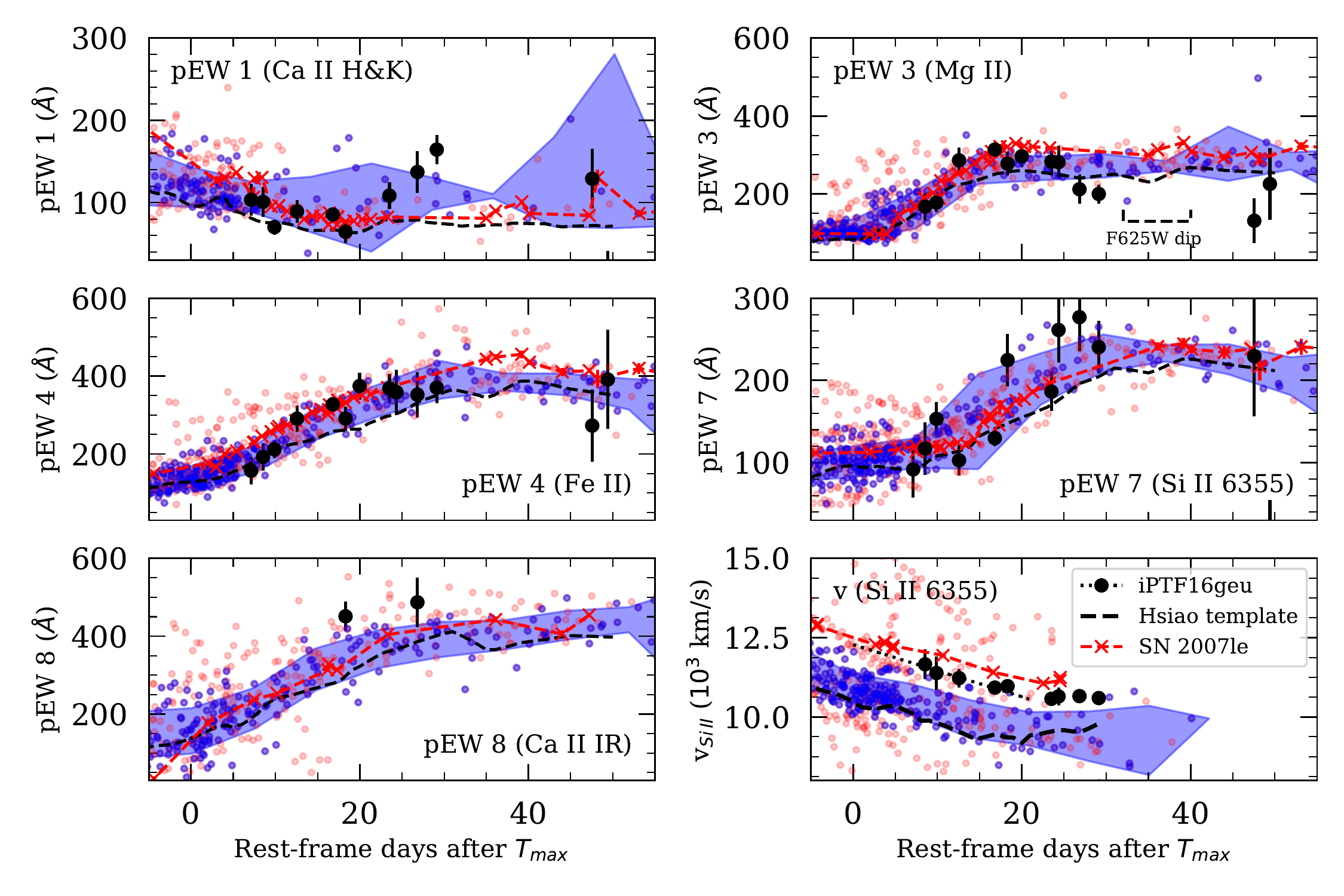}
	\caption{Time evolution of the pEWs for features 1,3,4,7 and 8 together with the \ion{Si}{ii} expansion velocity evolution (bottom right panel). Black circles are the measured pEWs after lens- and host galaxy subtraction and de-reddening. Transparent points and shaded bands are individual measurements and the binned mean ($\pm 1\sigma$) of low-redshift SNe from \citet{folatelli2013}, for all (red) and "Normal" (blue) SNe~Ia in their sample. The black and red dashed lines shows the pEW and velocity evolution of the Hsiao template and SN\,2007le, respectively.
	\label{fig:pews}}
\end{figure*}

We measure the \ion{Si}{ii}~$\lambda$6355 line expansion velocities for all our spectra. Using data between +7 and +20 days after maximum (although the clear identification with \ion{Si}{ii}~$\lambda$6355 is only valid until day $\sim$+10), we find a linear slope of the \ion{Si}{ii} expansion velocity, $\dot{v} = -82 \pm 13$ \kmsd, which is slower than what \citet{Cano2018} reports for the same data ($\dot{v} = -110\pm 10$ \kmsd, see Fig.~\ref{fig:pews}). Hence, the velocity gradient for \geu is more comparable to the normal sub-class in \citet{folatelli2013} ($\dot{v} = -86 \pm 14$ \kmsd), rather than the high-velocity gradient subclass. 
However, from the linear fit we extrapolate the velocity at $t_{B,max}$ to be $v_{B,max} = 12100 \pm 220$ \kms, which would make \geu a high-velocity \snia \citep[following the definitions in][]{Wang2009,folatelli2013}. We note that our velocities are systematically higher (by $\sim 400$ km/s) than in \citet{Cano2018}. 

Turning to the pEW measurements, we do not see any significant deviations from the time-evolution of SNe~Ia in \citet{folatelli2013}.  In Figure \ref{fig:pews}, the black points show the pEW measurements of \geu compared to all (red points) or "normal" (blue points and blue shaded region) SNe~Ia in  \citet{folatelli2013}. However, we note that some measurements are outside the 1$\sigma$ range, e.g. late time pEW1 (likely due to improper lens and host galaxy removal), pEW3 and pEW7 (where the telluric corrections are imperfect). 
\begin{table*}
	\centering
	\caption{%
		Measured pseudo-equivalent widths (pEW) and \ion{Si}{ii}~$\lambda$6355 line velocities for iPTF16geu. Last two colums list the total \ion{Na}{id} equivalent widths for the lens and host galaxies from our highest-resollution spectra.}
		\label{tab:pew_vel}
	\begin{tabular}{r | c c c c c c | c c}
\hline\hline
Phase & pEW1  & pEW3  & pEW4 & pEW7  & pEW8  & $v_{\rm \ion{Si}{ii}\lambda 6355}$ & Lens \ion{Na}{id} EW & Host \ion{Na}{id} EW  \\
(days) & (\AA) & (\AA) & (\AA) & (\AA) & (\AA) & ($10^{3}$ km/s) & (\AA) & (\AA)  \\  
\hline
7.4  & 103 (19) &          & 157 (35) & 92 (34)  &         &               &         &\\
8.8  & 101 (18) & 167 (35) & 192 (35) & 117 (32) &	        & 11.66 (0.47)  &         & \\
10.2  &  70 (9)  & 177 (25) & 212 (23) & 153 (20) &         & 11.38 (0.53)  &         & \\
12.8 &  75 (15) & 256 (38) & 271 (39) & 80 (24)  &         & 11.22 (0.97)  &         & \\
17.1 &  79 (12) & 295 (26) & 320 (26) & 95 (15)  &         & 11.07 (0.73)  &         & \\
18.6 &  64 (14) & 277 (31) & 291 (30) & 224 (31) & 451 (38) & 10.85 (0.36) & 2.4 (0.1) & 3.9 (0.1)\\
23.8 & 108 (16) & 283 (28) & 368 (36) & 186 (24) &         &               & 2.3 (0.2) & 3.5 (0.2) \\ 
27.1 & 137 (26) & 212 (37) & 352 (57) & 277 (42) & 487 (63) &              & 2.3 (0.1) & 3.3 (0.1) \\ 
29.4 & 164 (18) & 200 (25) & 370 (40) & 241 (32) &         &               &         &\\
47.9 & 129 (37) & 131 (57) & 273 (93) & 230 (73) &         &               &         &\\
49.7 & 13 (28) & 226 (92)  & 391 (127) &         &         &               &         &\\
\hline\hline
\end{tabular}
\end{table*}

\section{Time-delay measurements from resolved spectra}\label{sec:timedelay}

In our single epoch of HST spectroscopy, we can resolve two spectra: one spectrum corresponding to Image 2 and one spectrum corresponding to Images 1, 3 and 4. However, Images 3 and 4 are subdominant, contributing with 7\% and 5\% to the total flux in $F814W$ at this epoch, respectively. Hence, we will treat this spectrum as stemming from Image 1. 
By comparing the time evolution of spectral features between the spectra, we can in  principle provide an independent measurement of the time delays between the SN images. However, in our case we are limited by the coarse spectral resolution. To complicate things further, at the epoch (+29 days) the \ion{Si}{ii} absorption feature is no longer well-defined, displaying two or more local minima.

We construct a simple $\chi^2$-statistic, fitting Hsiao template spectra \citep{hsiao2007} at different phases to the resolved spectra. We simultaneously fit for residual lens and host galaxy contamination using our template spectrum, and subtract a fraction of that flux ($0.80 \pm 0.03$ and $0.19 \pm 0.02$ for Images 1 and 2, respectively) from the resolved SN spectra, so that the subtracted SN spectra and Hsiao templates at each phase match the host subtracted $F814W$ photometry. 

The best fit phases for the Hsiao templates are +30.0$^{+2.5}_{-2.2}$ days for Image 1 and +31.0$^{+3.8}_{-3.5}$ days for Image 2. Assuming a SN stretch, $s=1.0$, this corresponds to $T_{\rm max,1} = 57652.0$ and $T_{\rm max,2} = 57650.6$, which corresponds to an observed time-delay between images 1 and 2 of $1.4^{+4.6}_{-4.1}$ days. 
This is, to our knowledge, the first spectroscopic time-delay ever measured. The time delay is compatible, but less constraining, than the time-delay measurements from light-curve fits in \citet{dhawan2020}, where they find $T_{\rm max,1} = 57652.80 (\pm 0.33)$ and $T_{\rm max,2} = 57652.57 (\pm 0.99)$, which corresponds to an observed time-delay between images 1 and 2 of $-0.23 \pm 1.04$ days. 

The fits are weakly dependent on the choice of lens and host galaxy template (i.e. if we use the modelled or observed spectrum), and insensitive to fitting reddened or unreddened Hsiao template SEDs. We also tried using the SED template of SN\,2011fe \citep{Amanullah:2015bj}, which gives best fit phases $31.1^{+2.9}_{-3.3}$ and $32.8^{+2.9}_{-5.7}$ days for Images 1 and 2, corresponding to an observed time-delay, $\Delta t_{12}= 2.4^{+4.0}_{-8.9}$ days). 
\begin{figure}
	\centering
	\includegraphics[width=\columnwidth]{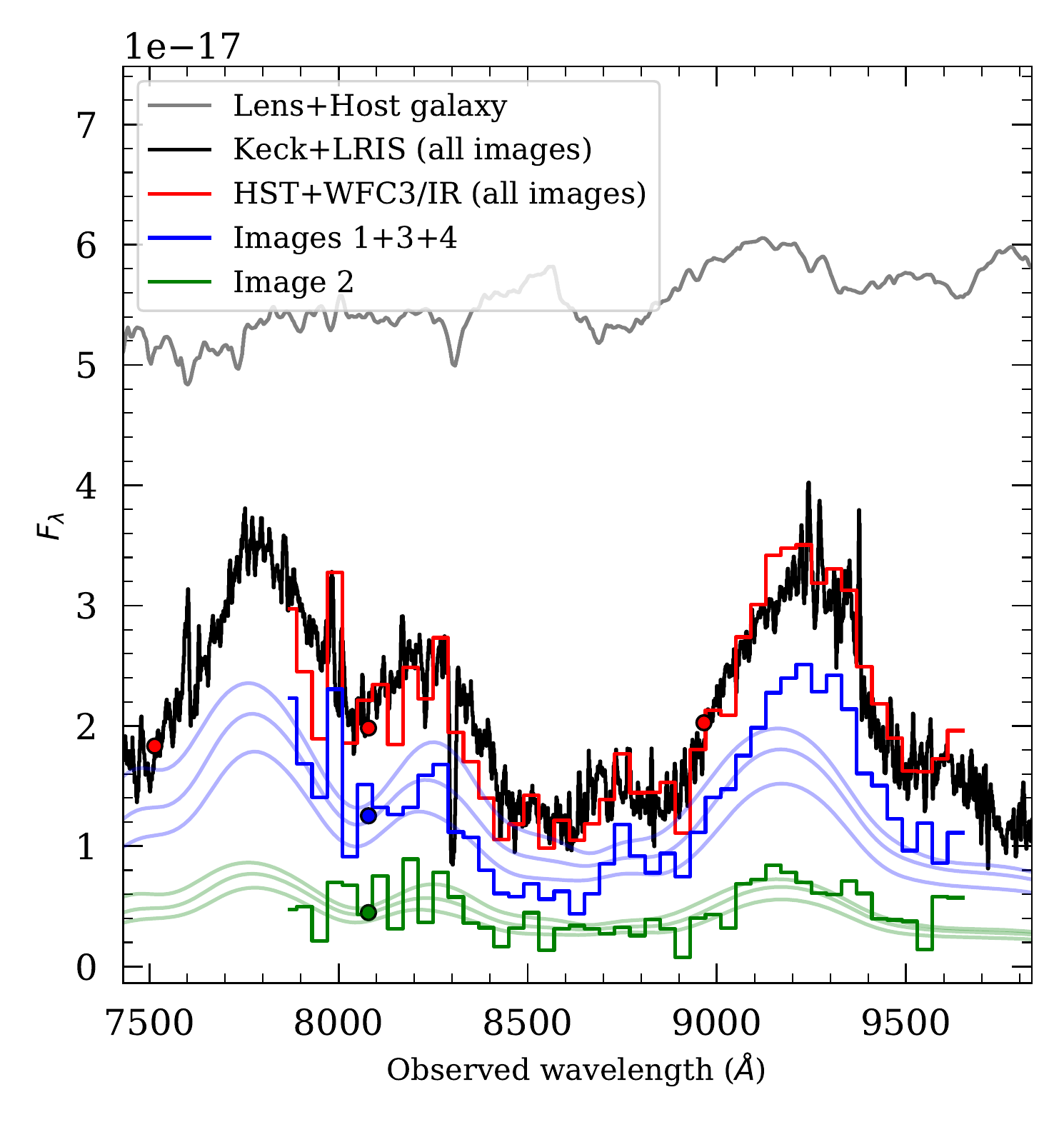}
	\includegraphics[width=\columnwidth]{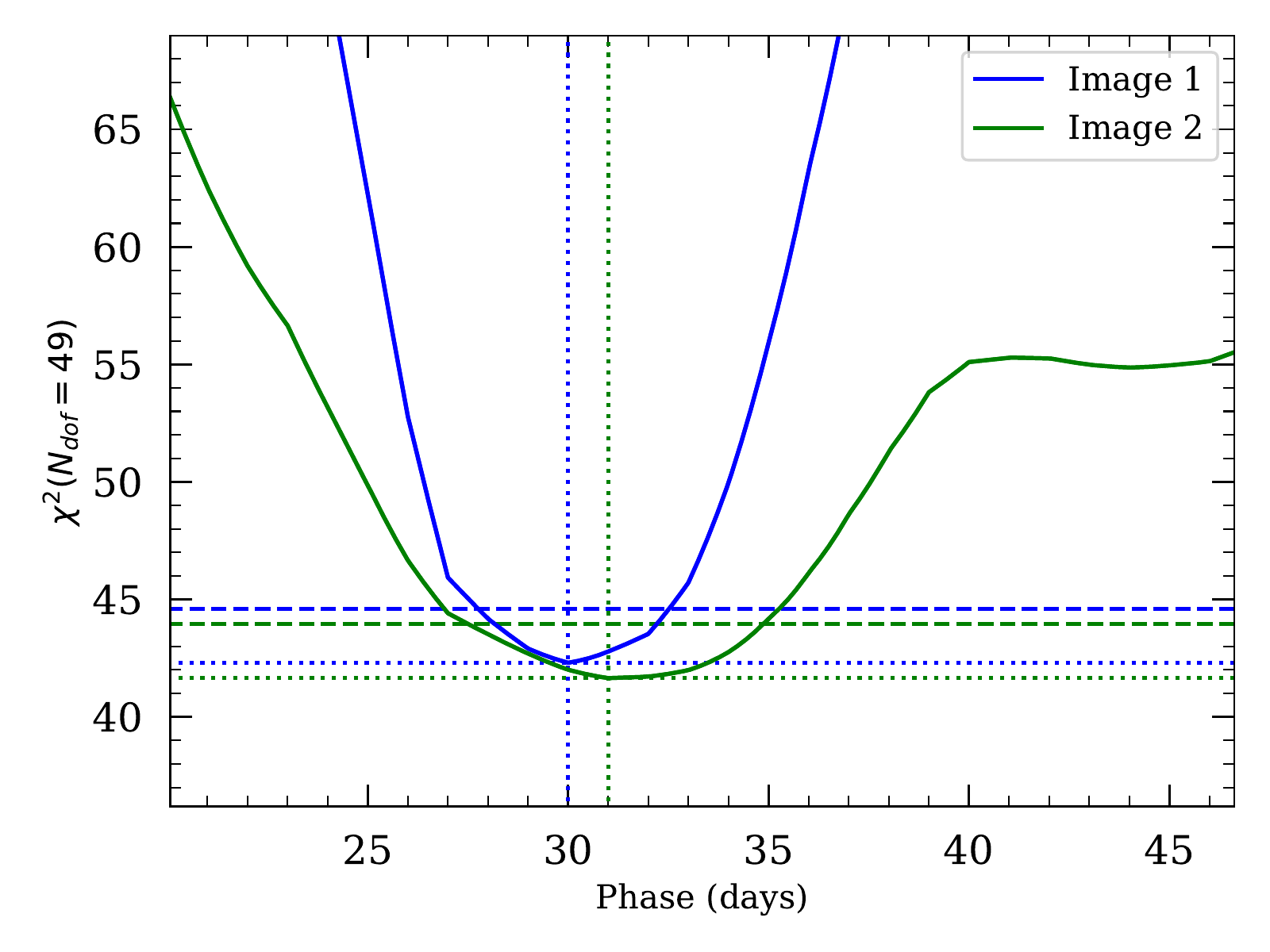}
	\caption{Top panel: Resolved HST spectra of Images 1, 3 and 4 (blue line) and Image 2 (green line). The summed spectra (red line) show very good agreement with the ground based Keck spectrum (black line), observed at the same time. Bottom panel: $\chi^{2}$ minima of the fitted phases for Images 1 (blue line) and Image 2 (green line).
    \label{fig:hstspec}}
\end{figure}

\section{Discussion}\label{sec:discussion}
Much attention has recently been given to microlensing effects, especially how chromatic distortion of the supernova spectra affect light-curve and time-delay measurements \citep{suyu2020}. While the light-curve analysis and lens modelling indicate that Image 1 (and possibly Image 2) are affected by microlensing, we do not see any clear spectroscopic signs of chromatic effects.

We do note a small dip in the $F625W$ light-curve of Image 1 ($\Delta m \sim 0.3$ mag, also seen in the summed photometry in Fig.~\ref{fig:lc}) around 50 days after peak brightness. This dip is only seen for Image 1 in two $F625W$ epochs (2016 Nov. 10 and Nov. 15), and is not seen for the other images nor in the other HST filters. Unfortunately, there are no spectroscopic observations during this dip. However, we do see a small decrease in pEW3 just before the onset of the dip.

While the pEW evolution could be a useful tool to detect or constrain chromatic lensing effects, it is difficult to quantify. For example, the pEWs are at all times consistent with the binned mean of the {\em sample} of normal \sneia in \citet{folatelli2013}, while if we compare the pEW evolution to an {\em individual} SN, the deviations as a function of wavelength and time can be larger or smaller depending on which SN we choose. Using SNID \citep{Blondin2007} to cross-correlate the \geu spectra with a library of well studied \sneia, SN\,2007le \citep{simon2009} appears among the top matches. In Figure~\ref{fig:pews}, we show the pEWs and \ion{Si}{ii} expansion velocity as function of time for \geu (black symbols) which closely follows the values for SN\,2007le (crosses and red dashed lines).

Turning to the strong \ion{Na}{id} absorption features in the host galaxy of \geu,
it is interesting to note that the deepest absorption feature is the most redshifted, placing \geu in the "blueshifted subclass" as defined in \citet{sternberg2011}. \citet{phillips2013} studied a large sample of \sneia with high-resolution spectra, and found that all events with anomalously large \ion{Na}{id} column densities (in comparison to the derived dust extinction from their colors) belonged to this "blueshifted subclass".
We also see a significant \emph{decrease} of the total Na ID EW in our highest resolution spectra: the \ion{Na}{id} EW goes from 3.9 to 3.3\,\AA\,between VLT+XSHOOTER epochs at +18 and +27 days ($EW=3.5$\,\AA\,in the Keck+DEIMOS spectrum at +24 days).

These facts also make a strong link to SN\,2007le \citep{simon2009} which also showed strong, blueshifted and time-variable \ion{Na}{id} absorption ($EW\sim 1.6$ \AA) but was not highly reddened ($E(B-V) = 0.27$ mag). It has been speculated that these \sneia, with strong, blueshifted absorption may belong to a distinct sub-population of SNe: having systematically higher ejecta velocities and redder colors at maximum brightness, preferentially residing in late-type galaxies \citep{foley2012b,maguire2013}. High-resolution spectroscopy of gravitationally lensed SNe thus offer us a way to study the progenitor systems and the explosion properties of high-redshift \sneia.

\section{Summary and outlook}\label{sec:summary}
In this paper, we have analysed the spectra of the lensed SN Ia iPTF16geu.
Using high dispersion spectra, we fit the lens galaxy line-of-sight velocity dispersion, $\sigma = 129 \pm 4$ km/s at a redshift $z_{\rm lens}=0.2163$. This value is lower than the previous estimate and compatible with the lens model in \citet{mortsell2019}. Knowing the velocity dispersion of the lens breaks degeneracies in the lens modelling, e.g. between the slope of the lens mass distribution and $H_0$ \citep{jee2019, mortsell2019}.

Using high angular resolution slitless spectroscopy we derive a spectroscopic time delay between the two brightest images (Image 1 and 2). Future spatially resolved spectra of lensed supernova discovered early by e.g. ZTF and LSST could yield independent time-delay measurements, with precision comparable to light-curve fits. Near maximum light, SNe Ia typically have expansion velocity gradients $\sim 100-250$ \kmsd. Provided spatially resolved spectra of the SN images, the time delays can be measured to roughly one day precision (given that e.g. the expansion velocities of absorption lines of the SN is observed at early phases, when the expansion velocity gradient is higher).
Since the typical expansion velocity decreases as $v_{\rm exp}(t) \propto t^{-0.22}$ \citep{piro2014}, spectra at even earlier phases would allow time-delays measurements with precision better than one day.

\section*{Acknowledgements}
We thank E. Zackrisson and A. Adamo for helpful discussions.
We thank T. Brink, R. Bruch, Cooper, S. Goldwasser, A. Ho, I. Irani, B. Jain, Y. Sharma, A. Tzanidakis, Q. Ye and W. Zheng for performing observations.

A.~G. acknowledges support from the Swedish National Space Agency and the Swedish Research Council. R.~L. is supported by a Marie Sk\l{}odowska-Curie Individual Fellowship within the Horizon 2020 European Union (EU) Framework Programme for Research and Innovation (H2020-MSCA-IF-2017-794467). A.~S.~C. acknowledges support from the G.R.E.A.T research environment, funded by {\em Vetenskapsr\aa det},  the Swedish Research Council.

The Intermediate Palomar Transient Factory project is a scientific collaboration among the California Institute of Technology, Los Alamos National Laboratory, the University of Wisconsin, Milwaukee (USA), the Oskar Klein Center at Stockholm University (Sweden), the Weizmann Institute of Science, the TANGO Program of the University System of Taiwan, and the Kavli Institute for the Physics and Mathematics of the Universe. 

Some of the data presented herein were obtained at the W.M. Keck Observatory, which is operated as a scientific partnership among the California Institute of Technology, the University of California and the National Aeronautics and Space Administration. The Observatory was made possible by the generous financial support of the W.M. Keck Foundation. The authors wish to recognize and acknowledge the very significant cultural role and reverence that the summit of Mauna Kea has always had within the indigenous Hawaiian community. We are most fortunate to have the opportunity to con- duct observations from this mountain. 

These results made use of the Lowell Discovery Telescope at Lowell Observatory. Lowell is a private, non-profit institution dedicated to astrophysical research and public appreciation of astronomy and operates the DCT in partnership with Boston University, the University of Maryland, the University of Toledo, Northern Arizona University and Yale University. The upgrade of the DeVeny optical spectrograph has been funded by a generous grant from John and Ginger Giovale and by a grant from the Mt. Cuba Astronomical Foundation.





\bibliographystyle{mnras}
\bibliography{geu_spectroscopy} 







\bsp	
\label{lastpage}
\end{document}